# Electrical and Thermal Transport in Metallic Single-Wall Carbon Nanotubes on Insulating Substrates

Eric Pop,[1,2,*] David A. Mann,[1] Kenneth E. Goodson[2] and Hongjie Dai[1]

[1]Laboratory for Advanced Materials and Department of Chemistry
[2]Thermal Sciences Department, Mechanical Engineering
Stanford University, Stanford CA 94305, U.S.A.
[*]Currently with Intel Corporation, Santa Clara, CA 94054

We analyze transport in metallic single-wall carbon nanotubes (SWNTs) on insulating substrates over the bias range up to electrical breakdown in air. To account for Joule self-heating, a temperature-dependent Landauer model for electrical transport is coupled with the heat conduction equation along the nanotube. The electrical breakdown voltage of SWNTs in air is found to scale linearly with their length, approximately as 5 V/μm; we use this to deduce a thermal conductance between SWNT and substrate $g \approx 0.17 \pm 0.03$ WK$^{-1}$m$^{-1}$ per tube length, which appears limited by the SWNT-substrate interface rather than the thermal properties of the substrate itself. We examine the phonon scattering mechanisms limiting electron transport, and find the strong temperature dependence of the optical phonon absorption rate to have a remarkable influence on the electrical resistance of micron-length nanotubes. Further analysis reveals that unlike in typical metals, electrons are responsible for less than 15% of the total thermal conductivity of metallic nanotubes around room temperature, and this contribution decreases at high bias or higher temperatures. For interconnect applications of metallic SWNTs, significant self-heating may be avoided if power densities are limited below 5 μW/μm, or if the SWNT-surrounding thermal interface is optimized.

Contact: epop@stanfordalumni.org, hdai@stanford.edu

## I. INTRODUCTION

Single-wall carbon nanotubes (SWNTs) are cylinders formed by a sheet of hexagonally arranged carbon atoms (graphene) wrapped into a nanometer-diameter tube [1, 2]. These molecular wires have attracted considerable scientific and engineering interest due to their outstanding electrical and thermal transport properties. Depending on their wrapping (chiral) angle, SWNTs exhibit either semiconducting or metallic behavior [3]. Within integrated circuits, semiconducting SWNTs can be used for transistor device applications [4], while metallic SWNTs have been proposed as advanced interconnects [5]. Compared to typical copper interconnects, SWNTs can carry up to two orders of magnitude higher current densities (~ $10^9$ A/cm$^2$) and are insensitive to electromigration.

Several recent studies have analyzed the promise of metallic SWNTs as circuit elements [6-9]. However, little prior work has investigated the influence of temperature on their electrical properties, an essential step in understanding their behavior within integrated circuits. Despite their high thermal conductivity, the thermal *conductance* of carbon nanotubes is relatively low owing to their small diameter and thermal boundary resistance with the environment [10, 11]. This implies that significant self-heating of SWNTs occurs under high current conditions [12, 13], a fact often overlooked in earlier studies of high bias transport in SWNTs on insulating substrates [14-16].

In this study we focus on electrical and thermal transport in metallic SWNTs on insulating substrates, under a wide range of temperature and bias conditions. We use in-air Joule breakdown of SWNTs to empirically extract the value of their thermal conductance to the insulating substrate. Furthermore, we describe an electro-thermally self-consistent transport model suitable for circuit simulation, and compare our analysis with experimental measurements. This also represents the first comprehensive description of a temperature-dependent Landauer approach to one-dimensional transport. We find that the low-bias electrical resistance of metallic SWNTs with lengths relevant to interconnect applications (microns) is affected by a strong temperature dependence of the optical phonon (OP) absorption length above 250 K. Consequently, we also thoroughly explore the temperature dependence of all relevant electron scattering mechanisms. Our results have significant implications for the viability of SWNT-based interconnects, and the proposed model covers a wide range of voltages and temperatures of practical interest.

## II. JOULE HEATING AND BREAKDOWN IN AIR

In order to study the influence of Joule self-heating on electrical transport in SWNTs on insulating substrates, the heat loss coefficient from the tube into the substrate must be known (the *g* term in Fig. 1 and Eq. 1). This can only be estimated if both the temperature and the electrical power dissipated in the SWNT are simultaneously available. The latter is known from the electrical *I-V* measurement, but the temperature profile of an individual nanotube is very difficult to obtain experimentally in a quantitative manner [17]. However, the temperature under which SWNTs break down by oxidation when heated and exposed to air is relatively well established from thermogravimetric analysis (TGA) experiments [18-20]. Here, we exploit this mechanism to gain the additional insight into the SWNT electrical and thermal properties, and the high temperature necessary for in-air breakdown is provided by Joule self-heating at high applied bias. We assume breakdown occurs when the middle of the tube (the point of highest temperature) reaches $T_{BD}$ = 600 °C = 873 K, in accord with the temperature suggested by TGA experiments.

Figure 1 shows a schematic of the SWNT-on-substrate layout used in the experiments and the model presented here. The heat conduction equation along the length of a current-carrying SWNT, including heat generation from Joule self-heating and heat loss to the substrate is [21]:

$$A\nabla(k\nabla T) + p' - g(T - T_0) = 0 \quad (1)$$

where $A = \pi db$ is the cross-sectional area ($b \approx 0.34$ nm the tube wall thickness), $k$ is the SWNT thermal conductivity [11], $p'$ is the Joule heating rate per unit length, and $g$ is the net heat loss rate to the substrate per unit length. An analytic expression for the temperature profile $T(x)$ along the SWNT can first be obtained if we make the simplifying assumption of uniform heat generation along the tube, $p' \approx I^2(R-R_C)/L$ where $R$ is the total electrical resistance of the nanotube and $R_C$ is the combined electrical resistance of the two contacts. This is reasonable, given the relatively flat expected temperature profiles, especially in longer tubes [17, 21] and yields:

$$T(x) = T_0 + \frac{p'}{g}\left[1 - \frac{\cosh(x/L_H)}{\cosh(L/2L_H)}\right] \quad (2)$$

for $-L/2 < x < L/2$, where $T_0$ is the temperature of the contacts at the two ends and $L_H = (kA/g)^{1/2}$ is the characteristic thermal healing length along the SWNT [22]. The prefactor $p'/g$ is simply the peak temperature rise in the middle of the tube when $L >> L_H$, i.e. for SWNTs several microns long ($L_H \approx 0.2$ µm, as discussed in Section V below). Breakdown occurs when the maximum temperature of the tube reaches the value of the breakdown temperature ($T_0 + p'/g = T_{BD}$), which allows the extraction of a simple expression for the breakdown voltage of long SWNTs:

$$V_{BD} = gL(T_{BD} - T_0)/I_{BD} \quad (3)$$

where we have approximated $p' = I_{BD}V_{BD}/L$ as the Joule power dissipated per unit length at breakdown. The breakdown current of metallic SWNTs longer than about 1 µm with good electrical contacts is well-known to be generally very close to $I_{BD} \approx 20$ µA when laying on insulating substrates, as observed by previous researchers [14-16] as well as in the course of this work. Consequently, the expression in Eq. 3 does well to reproduce the linear relationship between in-air breakdown voltage and SWNT length observed empirically, in Fig. 2. The approximation is less accurate for SWNTs shorter than 1 μm, when a significant portion of the Joule heat is generated *and* dissipated at the contacts rather than along the length of the tube itself. For simplicity, neither the thermal contact resistance nor the heat dissipation at the contacts was taken into account in the expression of Eq. 3 above. Using the finite-element model described in the later sections of this work we have found that the breakdown voltage of SWNTs shorter than 1 μm is consistent with a contact thermal resistance $\mathcal{R}_{C,Th} \approx 5 \times 10^6$ K/W [11, 21], however more detailed future efforts must be made to investigate electro-thermal transport in such short metallic SWNTs. In addition, we note that when the electrical resistance of the contacts ($R_C$) is significant, the breakdown voltage in Eq. 3 is increased by approximately $I_{BD}R_C$ as

$$V_{BD} = gL(T_{BD} - T_0)/I_{BD} + I_{BD}R_C \tag{4}$$

and hence the breakdown *power* is simply the expression above multiplied by the breakdown current.

Empirically, for metallic SWNTs with low $R_C$ and $L > 1$ μm, we find the in-air breakdown voltage ($V_{BD}$) scales linearly with the length of the tubes approximately as 5 V/μm, indicated by the dashed line in Fig. 2. This suggests that longer tubes benefit from better heat sinking into the substrate along their length, the net value of this thermal conductance being proportional to the length $L$. In other words, metallic SWNTs of 3 μm length will break down in air at about 15 V (at typical room temperature and pressure), when the electrical contact resistance is relatively low, as achievable with Pt or Pd contacts, which is exactly what was found in Ref. [14]. In separate experiments we have noted that breakdown voltages in an argon ambient were nearly twice as large [23]. Assuming an in-air breakdown temperature $T_{BD} = 600$ °C = 873 K as previously mentioned, we extract a net thermal conductance to substrate $g \approx 0.17$ WK$^{-1}$m$^{-1}$ per tube length by fitting Eq. 3 to the empirical data in Fig. 2. This is the value used in all calculations for the remainder of this work, unless stated otherwise. Given the maximum reported range of breakdown temperatures for SWNTs exposed to air ($500 < T_{BD} < 700$ °C) [18-20], the thermal conductance to substrate is expected to fall in the $0.14 < g < 0.20$ WK$^{-1}$m$^{-1}$ range. For the approximate contact area between nanotube and substrate ($A_C \approx Ld$), this is equivalent to a thermal resistance per unit area ($\rho = d/g$) between $1-2 \times 10^{-8}$ m$^2$KW$^{-1}$ for the span of thermal conductances ($g$ above) and SWNT diameters ($d \sim 2-3$ nm) relevant here. In units of thermal conductance per unit area this corresponds to $0.5-1 \times 10^8$ WK$^{-1}$m$^{-2}$, which falls well within the range of interface thermal conductances measured between various solids [24-26], suggesting it is the very narrow heat flow constriction at the nanotube-substrate interface which dominates heat sinking from the SWNT, rather than the material of the substrate itself.

At this point, it is instructive to compare these numbers with the estimated thermal spreading conductance from a heated narrow cylinder resting on a thermally insulating planar slab, as schematically drawn in Fig. 1. One must be careful with the choice of thermal resistance model used to estimate this heat loss. For instance, the problem at hand (very narrow cylinder of diameter $d \ll t_{ox}$) is quite different from the heated rectangular nanowire described by Durkan *et al.* [27] (rectangular nanowire of width $w \gg t_{ox}$), which is not applicable here. The heat loss in the latter case is essentially one-dimensional, whereas heat spreading from the very narrow circular SWNT is two-dimensional (see Fig. 1b). Following a standard reference [28] for this type of geometry, the conductance per unit nanotube length owed to the insulator (here, silicon dioxide) alone can be estimated as:

$$g_{ox} = \frac{\pi k_{ox}}{\ln\left(\frac{8 t_{ox}}{\pi d}\right)} \qquad (5)$$

where $k_{ox}$ and $t_{ox}$ are the thermal conductivity and thickness of the insulator, respectively. This yields estimates of $g_{ox} \approx 1.7$ and $1.0$ WK$^{-1}$m$^{-1}$ for $t_{ox} \approx 10$ and 67 nm in the work of Javey *et al.* [14], and $g_{ox} \approx 0.8$ WK$^{-1}$m$^{-1}$ for the thicker $t_{ox} \approx 200$ nm in the work of Park *et al.* [15], assuming $k_{ox} \approx 1.4$ Wm$^{-1}$K$^{-1}$ (for thermally grown silicon dioxide) and a typical SWNT diameter $d \approx 2$ nm. Hence, the *total* value of the nanotube-substrate conductance $g \approx 0.17$ WK$^{-1}$m$^{-1}$ inferred above from our burning temperature arguments is 5–10 times lower than the conductance owed to these insulating substrates alone. This also strongly indicates that the atomically narrow heat flow constriction at the nanotube-substrate interface [29, 30] is the factor limiting heat sinking from the SWNT, rather than the material or thickness of the insulating substrate itself. The latter accounts for only 10–20% of the total thermal resistance from the SWNT along its length, while any additional spreading thermal resistance introduced by the remaining thickness of the typical silicon wafer is almost two orders of magnitude lower, and therefore negligible.

The recent results of Maune *et al.* [31] (which we became aware of during the preparation of this manuscript) also support the conclusions we draw above. They infer a thermal conductance per unit length about 0.26 WK$^{-1}$m$^{-1}$ between SWNTs and a sapphire substrate, the slightly larger extracted value (compared to ours) possibly influenced by higher electrical resistance of their Au/Cr contacts (see Eq. 4). Nevertheless, this value is also very much in the range of typical solid-solid interface thermal conductance, and *inconsistent* with the higher thermal conductivity of the sapphire substrate vs. the silicon dioxide substrates considered in this work. Once again, the SWNT-substrate interface, rather than the thermal properties of the specific substrate, appears to dominate heat dissipation from the nanotube.

### III. ELECTRICAL TRANSPORT

In order to understand the electrical behavior of the metallic SWNT over a wide range of temperatures and biases, we develop a straightforward, yet fully coupled electro-thermal transport model. Figure 1 shows the SWNT layout considered in this work and Fig. 3 illustrates the *I-V* characteristics of a typical 3 μm long metallic tube, up to breakdown by oxidation in air. The temperature dependence of the nanotube resistance is obtained through the temperature dependence of the electron scattering mean free paths (MFPs) with acoustic (AC) and optical (OP) phonons [12]. The temperature profile $T(x)$ along the SWNT is computed consistently with the power dissipated by Joule heating from Eq. 1 above. The total electrical resistance is written similarly to the Landauer-Büttiker approach [32, 33], but summing over the series contribution of individual SWNT segments of length $dx$, each at a temperature $T(x)$ as

$$R(V,T) = R_C + \frac{h}{4q^2}\left[1 + \int_{-L/2}^{L/2} \frac{dx}{\lambda_{eff}(V,T(x))}\right] \quad (6)$$

where $R_C$ is the combined electrical resistance of the two contacts, not including the quantum contact resistance $h/4q^2$ (arising from the mismatch of conduction channels in the SWNT and the macroscopic electrodes) which is accounted for in the second term. The factor of 4 accounts for the four parallel 1-D conduction channels of a SWNT, owed to spin degeneracy and the band degeneracy of graphene [1, 2]. The integrand with its prefactor are interpreted as the resistance of a nanotube portion of length $dx$, i.e. $dR = (h/4q^2)(dx/\lambda_{eff})$, such that the Joule heating rate per unit length in Eq. 1 is given by

$$p' = I^2 \frac{dR}{dx} = I^2 \frac{h}{4q^2} \frac{1}{\lambda_{eff}} \quad (7)$$

which is position, field and temperature dependent through $\lambda_{eff}$. This total, effective electron MFP is obtained from Matthiessen's approximation [34] as

$$\lambda_{eff} = \left(\lambda_{AC}^{-1} + \lambda_{OP,ems}^{-1} + \lambda_{OP,abs}^{-1}\right)^{-1} \quad (8)$$

which includes elastic electron scattering with AC phonons, and inelastic electron scattering by OP emission and absorption. We note that the latter has been neglected in much previous work due to the large OP energy in SWNTs ($\hbar\omega_{OP} \approx 0.16$–$0.20$ eV) and their low occupation near room temperature [14-16]. The electron-phonon scattering MFPs in metallic SWNTs can be expressed as $\lambda = v_F\tau$, where $v_F \approx 8 \times 10^5$ m/s is the electron Fermi velocity [1] and $\tau$ is the respective electron-phonon scattering time. Around room temperature and above the AC phonon modes are thermally occupied ($k_BT \gg \hbar\omega_{AC}$), and their scattering rate ($1/\tau_{AC}$) is linearly proportional with temperature [15]. Consequently, we can express the (elastic) AC scattering MFP at various temperatures as a function of the MFP at room temperature:

$$\lambda_{AC} = \lambda_{AC,300}\left(\frac{300}{T}\right), \quad (9)$$

where $\lambda_{AC,300} \approx 1600$ nm is the AC scattering length at 300 K. Owing to their high energy, the OP phonon occupation and (inelastic) scattering rates must be considered more carefully. The electron scattering rate with OP phonons can be expressed through Fermi's Golden Rule [35] as proportional to

$$\frac{1}{\tau_{OP}} \propto \left(N_{OP} + \frac{1}{2} \mp \frac{1}{2}\right) D(E \pm \hbar\omega_{OP}) \tag{10}$$

where $N_{OP} = 1/[\exp(\hbar\omega_{OP}/k_B T) - 1]$ is the optical phonon occupation and $D(E)$ is the density of states for the final electron state after scattering. The upper and lower signs correspond to OP absorption and emission, respectively. Due to the large energy of optical phonons in SWNTs ($\hbar\omega_{OP} \approx 0.16$–$0.20$ eV $\gg k_B T$), the optical phonon occupation has often been assumed to be $N_{OP} \approx 0$ in previous work, which implies negligible absorption. This assumption is relaxed here and $N_{OP}$ is explicitly included. Furthermore, we note that for metallic tubes the electron density of states $D(E)$ is a constant independent of energy [36]. This is also a good approximation for gated semiconducting tubes biased strongly away from the band gap. Hence, from Eq. 10 and the preceding discussion, the OP phonon emission and absorption rates will scale with temperature simply as ratios of the phonon occupation terms. This is the key insight which simplifies our temperature-dependent model and renders this approach almost entirely analytical. We can now write the OP absorption length as [12]:

$$\lambda_{OP,abs}(T) = \lambda_{OP,300} \frac{N_{OP}(300) + 1}{N_{OP}(T)} \tag{11}$$

where $\lambda_{OP,300} \approx 15$ nm is the spontaneous OP *emission* length at 300 K. This short (15 nm) distance is the characteristic length scale for OP phonon emission by the hot electrons whose energy *exceeds* the OP emission threshold $\hbar\omega_{OP}$. However, this is *not* the average OP emission MFP, which is much longer when considering that not all electron energies exceed the OP threshold. Most electrons have to gain sufficient energy from the electric field before overcoming this threshold, a length scale captured by the first right-hand term in Eq. 13 below. We also note that OP emission can occur both after electrons gain sufficient energy from the electric field, and after an OP absorption event:

$$\lambda_{OP,ems} = \left(1/\lambda_{OP,ems}^{fld} + 1/\lambda_{OP,ems}^{abs}\right)^{-1}. \tag{12}$$

The former MFP can be written as

$$\lambda_{OP,ems}^{fld}(T) = \frac{(\hbar\omega_{OP} - k_B T)}{qV/L} + \frac{N_{OP}(300) + 1}{N_{OP}(T) + 1}\lambda_{OP,300} \tag{13}$$

where the first term estimates the distance electrons of average energy $k_B T$ must travel in the electric field $F = V/L$ to reach the OP emission threshold energy [12, 15], and the second term represents the temperature dependence of the OP emission length beyond this threshold. The OP emission MFP after an absorption event is obtained from Eq. 13 by replacing the first term with the OP absorption length of Eq. 11:

$$\lambda_{OP,ems}^{abs}(T) = \lambda_{OP,abs}(T) + \frac{N_{OP}(300)+1}{N_{OP}(T)+1}\lambda_{OP,300}. \tag{14}$$

This approach lets us express the temperature dependence of the relevant MFPs with respect to the acoustic and optical scattering lengths at 300 K. The relatively simple method works because the scattering lengths scale as ratios of the phonon occupation terms for metallic SWNTs, whose density of states is nearly constant, as mentioned above. We note that in the limit of very small OP occupation (below room temperature) our $N_{OP}$ approaches zero, OP absorption can be neglected and the MFPs estimated above reduce to those of Refs. [14-16].

## IV. COUPLED ELECTRO-THERMAL TRANSPORT

The temperature profile along a current-carrying SWNT depends on the Joule power dissipated, and hence on its resistance. In order to obtain the *I-V* curves, equations 1 and 6–16 are computed self-consistently along the length of the tube. We use a damped (over-relaxed) iterative approach, with the temperature and other quantities of the previous iteration being used as initial conditions for the next [37, 38]. This method aids convergence, in the sense that the temperature profile at iteration *j* is "damped" using the temperature at the previous step as

$$T_j = \alpha T_j + (1-\alpha T_{j-1}) \tag{15}$$

where the damping parameter is $0 < \alpha < 1$. Naturally, a choice of $\alpha$ too small can make the model too slow, while a value too close to unity may lead to convergence problems. We have found values in the range 0.2-0.4 to be suitable here. At the beginning of a new iteration the $T_j(x)$ profile is first obtained from the solution of the heat conduction equation (Eq. 1), but it is the damped value of $T_j(x)$ that is then used to compute the MFPs and consequently the power dissipation $p'(x)$ in the next iteration. The new distribution of power dissipation is again introduced as the input to the heat conduction equation, and this iterative approach continues until the temperature profiles $T_j$ and $T_{j-1}$ converge within 0.1 K of each other.

The current is simply $I = V/R$, where the electrical resistance *R* depends on temperature and bias as described above. For completeness, the choice of thermal conductivity model for metallic SWNTs incorporated in our numerical solution depends on their length and the absolute temperature as [11]

$$k(L,T) = \left[ 3.7\times10^{-7}T + 9.7\times10^{-10}T^2 + 9.3(1+0.5/L)T^{-2} \right]^{-1} \tag{16}$$

which has an approximately $1/T$ dependence above room temperature and a slightly steeper $1/T^2$ component at temperatures approaching the burning point [11]. However, since the dominant heat conduction pathway is down into the substrate, the results of this study are not very sensitive to thermal quantities that influence conduction *along* the tube (such as *k* or *d*). For SWNTs longer than 1 μm resting on insulating substrates, we find the thermal conductance to the substrate ($g \approx 0.17$ WK$^{-1}$m$^{-1}$ per tube length, found in Section II) plays a more important role in determining their thermal behavior. The coupled model described here is similar to that presented with less detail in Ref. [12], but including heat loss to the substrate and excluding the OP non-equilibrium effect. We have found the latter is not required to reproduce the electrical characteristics of SWNTs on insulating substrates, as have other recent studies based on solutions of the Boltzmann Transport Equation [13]. This does not rule out some degree of OP non-equilibrium in nanotubes on substrates [39], but it does imply that OP non-equilibrium appears much weaker in SWNTs on substrates compared to freely suspended SWNTs [12]. The nature and vibrational modes of the substrate itself are thought to play a significant role in reducing the OP non-equilibrium effect: solid substrates being most effective, followed by polyatomic gases, and finally by monatomic gases "touching" the tube [40]. In other words, the more external vibrational modes are available in contact with the SWNT, the shorter the OP phonon lifetimes of the SWNT (as they find additional decay pathways), and the higher the current-carrying capability of the nanotube. In principle, a choice of substrate with vibrational modes closely related (coupled) with those of the SWNT could be made, to provide the shortest OP phonon lifetimes and the possibility of enhanced current flow at high bias.

## V. CURRENT-VOLTAGE CHARACTERISTICS

The model is compared to *I-V* data taken on a 3 μm long and 2 nm diameter metallic SWNT in Fig. 3 [14]. We note the current saturation close to 20 μA, and the subsequent breakdown of this tube when its peak temperature (in air) reaches approximately 600 °C. For the 3 μm long SWNT this occurs when the applied bias is near 15 V, consistently with our discussion in Section II (about 5 V/μm scaling of the breakdown voltage with length). It is apparent that the current saturation and eventual breakdown of the SWNT are both thermally limited and activated effects, respectively. The temperature profile along the nanotube is plotted in the figure inset, at biases of 3, 6, 9, 12 and 15 V from bottom to top. Since the last bias in this list is very close to the breakdown point, the peak nanotube temperature can be seen approaching 873 K. The length scale over which the temperature profile flattens out away from the contacts is $L_H = (kA/g)^{1/2}$ from Eq. 2 (the characteristic thermal healing length), which is approximately 0.2 μm. This suggests that all nanotubes longer than about 1 μm (the length range considered in this study) have relatively flat temperature profiles under Joule heating at high bias [21]. Nevertheless, our model relies on a complete discretization of all equations previously mentioned along the length of the nanotube, and the MFPs and temperature profile are computed at each point. Regarding electrical transport at high bias, we ought to point out that more generally, depending on the nanotube diameter and applied field, additional (higher) electron conduction channels may come into play [14, 41], although this is not evident for the particular SWNT studied in Fig. 3.

We explore the sensitivity of the model to several parameters in Fig. 4, with the solid line corresponding to the same *I-V* as in Fig. 3, and displayed here for reference. Increasing the room temperature spontaneous OP emission MFP ($\lambda_{OP,300}$ in equations 11, 13 and 14) from 15 to 20 nm yields a higher saturation current, but too low an in-air breakdown voltage (vs. experiment), since the total dissipated power must be the same to yield the same temperature. Once again, we note that $\lambda_{OP,300}$ should *not* be mistaken for the average electron-OP scattering MFP, but rather it is only the length scale traveled by electrons with energies greater than the OP energy ($\hbar\omega_{OP} = 0.17$ eV) before OP emission occurs, which is why we are calling it the *spontaneous* OP emission MFP. Most electrons have energies below $\hbar\omega_{OP}$, needing to travel farther to gain additional energy from the electric field to exceed this threshold and emit OPs, as shown by the first term on the right hand side of Eq. 13. In addition, all OP scattering lengths are dependent on temperature, as captured by the factors multiplying $\lambda_{OP,300}$ in equations 11, 13 and 14, with the OP absorption MFP having the strongest temperature dependence (Eq. 11 and Fig. 7).

Another parameter varied in Fig. 4 is the OP energy itself, which is known to be in the range 0.16-0.20 eV, between the Brillouin zone-edge (K-point) and zone-center (Γ-point) optical phonon energy [15, 16, 39, 42, 43]. We used $\hbar\omega_{OP} = 0.17$ eV to match the experimental *I-V* characteristics in Fig. 3, but values of 0.16, 0.18 and 0.20 eV have also been commonly used in previous research. Undoubtedly, the "true" microscopic mechanism limiting electron transport at high fields involves electron scattering with a *range* of phonons *around* the K- and Γ-points of the Brillouin zone, this range being broader at higher energies, when electrons find themselves farther away from the conduction band minima. In the context of the simple yet physical model presented in our approach it is reasonable to carefully select a suitable average, and to treat $\hbar\omega_{OP}$ as a fitting parameter within the physically appropriate bounds mentioned above. It is also not unlikely that the SWNT-substrate interaction (particular to each SWNT chirality and substrate surface) could affect one of the dominant OP modes, thus enabling preferential electron scattering with the other.

The role of the substrate may ultimately be of most importance in determining the heat dissipation coefficient from the SWNT, $g = 0.17$ WK$^{-1}$m$^{-1}$ in Eq. 1 of our model. We have increased this to 0.3 WK$^{-1}$m$^{-1}$ to obtain the dotted line in Fig. 4, a value not out of the realm of possibility assuming a better thermal quality of the SWNT-substrate interface. This would significantly increase the saturation current of the metallic SWNT, as well as its breakdown voltage in air, as the nanotube remains at lower temperatures for the same dissipated Joule power. While the SWNT-substrate thermal interface (and constriction) resistance [29, 30] may not be completely eliminated, it could be engineered to yield heat dissipation coefficients of the order 1 WK$^{-1}$m$^{-1}$ when the SWNT is resting on a thin insulating layer, as discussed in Section II of this manuscript.

The hypothetical situation when heat dissipation from tube to substrate is essentially perfect ($g \to \infty$) is examined theoretically in Fig. 5, for several SWNT lengths longer than one micron. This suggests that perfectly *isothermal* nanotubes ($T = T_0$) could yield significantly higher currents than the approximately 20 µA routinely observed so far in experiments at high applied biases for this SWNT length range [14-16]. Hence, it is worthwhile to understand the effect of voltage ($V$), current ($I$), total power ($IV$) and power density ($IV/L$) on self-heating and the electrical characteristics of the nanotube. Figure 5 shows that for a *given bias voltage* (consider, e.g., 3 V) longer SWNTs show less effect of self-heating owing to lower power densities ($p' \approx IV/L$ with lower current and longer length) and better heat sinking into the substrate along their length. For a *given current level* (consider, e.g., 15 µA) the average temperature rise in the SWNT is found to be essentially the same at all three lengths considered in Fig. 5 (e.g., approximately 20 K average temperature rise at 15 µA), although this same current is obviously reached at voltages that scale roughly as the respective nanotube lengths. This also implies that constant-current is equivalent to constant power density ($IV/L$) for SWNTs on insulating substrates in this length range, when self-heating is properly taken into account. In other words, iso-power density lines drawn on Fig. 5 would be essentially horizontal, just like iso-current lines (lines not actually drawn to avoid figure clutter).

It is evident that self-heating of SWNTs in the length range $1 < L < 10$ µm is non-negligible under high bias and current conduction. This is due to the high power density ($p' \approx IV/L$) and the large thermal resistances involved. Further analyzing Fig. 5, we generally find that power densities greater than approximately 5 µW/µm lead to noticeable self-heating of SWNTs in this length range, a simple rule which could be used as a design guideline. The assessment is more difficult for very short ($L < 1$ µm) SWNTs, as more power is dissipated at the contacts, which also play a stronger role in heat-sinking. In the present study the primary focus is on the electrical and thermal transport *within* the SWNT, as they dominate the behavior of micron-length tubes and are better understood than the electrical and thermal transport at the contacts. Future studies must address power dissipation issues in very short metallic nanotubes at high-bias, providing a more detailed understanding of the electrical and thermal transport at the contacts.

## VI. MEAN FREE PATH AND ELECTRICAL RESISTANCE

In Fig. 6 we turn to study the *low-bias* SWNT electrical resistance. The experimental data (symbols) were taken on a 3 µm long metallic SWNT similar to the one in Fig. 3, over a wide range of ambient temperatures. We note that in this case the ambient temperature fully dictates the transport in the SWNT, since at low bias the tube is essentially isothermal, and in thermal equilibrium with its surroundings. In Fig. 6a the experimental data is compared with our transport model including and excluding the OP absorption mechanism. Here we find that, although previously neglected, OP absorption plays a non-negligible role even at moderate to high temperatures ($T > 250$ K) and this scattering mechanism ought to be included in future studies of SWNTs in this length range ($L > 1$ μm). This is explained by the strong temperature dependence of the OP absorption MFP (Eq. 11), as clearly evident in Fig. 7. The expected temperature dependence of the low-bias resistance for various SWNT lengths is plotted in Fig. 6b, assuming otherwise perfect electrical and thermal electrode contacts ($R_C = 0$). Longer tubes have an earlier onset of OP phonon absorption, as is expected given the long OP absorption MFP at low bias (Fig. 7). The temperature coefficient of resistance (TCR) of metallic SWNTs around room temperature is ≈ 0.0026 K$^{-1}$, which is very near that of 40 nm diameter copper vias [44]. We note that the low-bias resistance of the longer tubes is expected to have the strongest temperature dependence, as the OP absorption length plays a less negligible role.

The temperature dependence of the various MFPs included in our model is shown in Fig. 7, as predicted by equations 8–14. Figure 7a displays the MFPs in the *low-bias* regime (60 mV across a 3 μm tube, or 200 V/cm field), while Fig. 7b samples the *high-bias* regime (6 V across a 3 μm tube, or 20 kV/cm field). Note the different temperature range covered by the two plots, chosen to provide more information. The OP absorption length is about 10 μm at room temperature and has the strongest temperature dependence (decrease) of all MFPs, which explains its non-negligible role especially at higher temperatures. Acoustic phonon scattering dominates transport at low bias as might be expected, with a $1/T$ dependence of the MFP, i.e. a linear dependence of the resistance on temperature, as in Fig. 6a. However, at temperatures beyond approximately 250 K the strong decrease in the OP absorption MFP imparts a steeper decrease to the overall MFP (black solid line in Fig. 7a), explaining the faster up-turn in the temperature dependence of the low-bias resistance, as in Fig. 6a. The OP absorption and AC scattering MFPs are only dependent on temperature, not on the electric field, hence their trace in Fig. 7b is simply a continuation of that in Fig. 7a. The OP emission MFP, however, is significantly shorter at high bias, becoming the dominant scattering mechanism. We note once again that the total OP emission MFP (Eq. 12) varies in the range of 50-100 nm over the high temperature regime up to the nanotube burning point, and that the parameter $\lambda_{OP,300} = 15$ nm constitutes only part of this MFP contribution, the mean distance necessary to emit an OP phonon once an electron reaches the $\hbar\omega_{OP}$ energy (also highlighted in Ref. [15]). Moreover, the OP emission MFP is strongly field-dependent: in low, 200 V/cm field at 300 K this MFP is about 4.5 μm (Fig. 7a), whereas in a high field of 20 kV/cm (6 V across 3 μm tube) it is about 92 nm at 300 K, and 53 nm at 850 K (Fig. 7b). In a field of 50 kV/cm (15 V across 3 μm tube) at 850 K the OP emission MFP reaches down to 32 nm, which essentially corresponds to the burning point in the *I-V* curve of Fig. 3. At the same high temperature the OP absorption and AC scattering MFPs are estimated to be 138 nm and 565 nm, respectively (Fig. 7b). In practice, of course, the specific combination of field and temperature (at every point along the nanotube) is determined by a self-consistent solution of equations 1 and 6–16, as described in the earlier parts of this manuscript.

**VII. ELECTRICAL AND THERMAL CONTACT RESISTANCE**

The electrical contact resistance of our SWNTs is of the order 10-100 kΩ with the Pt electrodes and is included in our model through the adjustable parameter $R_C$, as in Eq. 5. Note that in our model $R_C$ represents the combined contribution of *both* contacts. This mainly affects the slope of the nanotube *I-V* curve at low bias, where it appears in series with the quantum contact resistance ($h/4q^2 \approx 6.5$ kΩ) and with the intrinsic phonon-limited resistance of the SWNT (Fig. 6 and Section VI). In addition, $R_C$ is not expected to be strongly temperature-dependent, as the contacts remain essentially isothermal with the large electrodes.

The thermal contact resistance $\mathcal{R}_{C,Th}$ at the SWNT-electrode junction is of the order $5-10 \times 10^6$ K/W [11], and is included in our model by introducing Neumann boundary conditions in the heat conduction equation from Eq. 1:

$$k_L A \frac{dT_L}{dx} = \frac{T_L - T_0}{\mathcal{R}_{C,Th}} \qquad (17)$$

where the subscript "*L*" denotes the point along the SWNT nearest the left electrode contact, and $T_0$ is the ambient temperature of the electrode (e.g. 300 K). The boundary condition at the right electrode (subscript "*R*") can be similarly written. This creates a (relatively small) temperature "slip" at the SWNT-electrode junction, proportional to the contact thermal resistance and to the local heat flux. This slip is apparent in the inset of Fig. 3, where the high-bias temperature just inside the SWNT edges ($T_L$ at $x = -1.5$ μm and $T_R$ at $x = 1.5$ μm) becomes higher than $T_0 = 300$ K, following Eq. 17. The numerical value of the contact thermal resistance, as mentioned above, is consistent with the area of the thermal contact and with the typical solid-solid thermal resistance per unit area ($\rho \sim 10^{-8}$ m$^2$KW$^{-1}$, also see Section II).

Furthermore, we use the Wiedemann-Franz Law [45-47] to estimate the relative contribution of electrons and phonons to heat conduction along the nanotube and at the contacts. In its classical form this states that the part of the thermal conductivity owed to electrons is proportional to the electrical conductivity $\sigma$, the absolute temperature, and the Lorenz constant $L_0 = 2.45 \times 10^{-8}$ WΩK$^{-2}$ as

$$k_e = L_0 T \sigma = L_0 T \frac{L}{RA} \qquad (18)$$

which can also be expressed with respect to the length *L*, cross-sectional area *A*, and the electrical resistance *R*. Figure 8 compares the thermal conductivity of Eq. 16 (dashed line) with the thermal conductivity due to electrons alone (solid lines) estimated from Eq. 18 and our previous analysis of the electrical resistance at low- and high-bias conditions (Figs. 6 and 7). We note that the dashed line model of Eq. 16 is based on a high-bias extraction method [11], which inherently minimizes the electronic contribution to the thermal conductivity. Hence, the dashed trend is essentially equivalent to the phonon thermal conductivity, this being sufficient to analyze self-heating at high-bias in SWNTs. The electronic contribution to heat conduction is more significant at low-bias, where the SWNT electrical resistance is much lower, but self-heating itself is negligible at low-bias. From Fig. 8, we estimate that the electronic contribution to the SWNT thermal conductivity is at most 440 Wm$^{-1}$K$^{-1}$ at low-bias (less than 15% of the phonon thermal conductivity) and at most 60 Wm$^{-1}$K$^{-1}$ at high-bias (less than 5% of the phonon thermal conductivity at high temperature, and even smaller over the rest of the temperature range). We ought to note that, strictly speaking, the Wiedemann-Franz Law is best applied when transport is dominated by elastic scattering, otherwise only providing an upper limit for the electronic thermal conductivity.

To obtain additional figures of merit, Fig. 9 summarizes the resistance components of a 2 μm long metallic SWNT with 2 μm long contacts. From Eq. 16 we can also write down a relationship for the thermal resistance owed to electrons derived from the electrical resistance as $\mathcal{R}_{Th} = R/(L_0 T)$. Once again, thermal conduction owed to electrons both along the nanotube *and* at the contacts appears negligible around room temperature and above, also in accord with Ref. [48]. This is the case even for metallic nanotubes, a somewhat unusual result when remembering that electrons are the carriers responsible for heat transport in all typical metals [45]. For SWNTs this can be understood owing to their extremely high phonon thermal conductivity, greater than 3000 Wm$^{-1}$K$^{-1}$ at room temperature [11]. The electronic contribution to the thermal conductivity of metallic SWNTs may become more significant at temperatures below 50 K, as the phonons are quenched out [48], but this temperature range is outside the scope of the current study.

**VIII. CONCLUSIONS**

This work represents a study of electrical and thermal transport in metallic SWNTs relevant for practical interconnect applications ($L > 1$ μm), over a wide range of applied biases (up to electrical breakdown in air) and temperatures (approximately 100–800 K). Electron scattering by optical phonon absorption was found to play a previously neglected role in the *low-bias* resistance of long SWNTs, while Joule self-heating must be taken into account for transport in nanotubes at high-bias. In addition, the breakdown voltage of microns-long SWNTs exposed to air was found to scale linearly with the nanotube length, as explained by heat-sinking into the substrate along the nanotube. The heat dissipation coefficient into the substrate ($g \approx 0.17 \pm 0.03$ WK$^{-1}$m$^{-1}$) appears to be very much limited by the nanoscale constriction at the SWNT-substrate interface, rather than the thermal conductivity of the substrate itself.

We thoroughly described an electro-thermal model for transport in metallic SWNTs, based on a temperature-dependent Landauer approach self-consistently coupled with the heat conduction equation.

The method is transparent, easy to implement, and should enable other researchers to reproduce a variety of SWNT transport data over a wide range of practical voltages and temperatures. The model has been validated against experimental data, and is also readily usable for circuit simulators or other design studies. It appears that thermal management and design of high-current carrying nanotubes will be of importance for future interconnect and device applications.

**LIST OF FIGURES**

**Figure 1:** Schematic of single-wall carbon nanotube (SWNT) on insulating substrate, with Pt electrical contacts (not drawn to scale). The two cross-sections are longitudinal **(a)** and transverse **(b)** to the direction of the nanotube. The arrows indicate heat spreading from the nanotube under Joule self-heating. The thermal conductance from SWNT- and contacts-to-substrate is dominated by their interface, with $g \approx$ 0.14–0.20 WK$^{-1}$m$^{-1}$ per unit nanotube length.

**Figure 2:** Experimentally measured breakdown voltages ($V_{BD}$) for metallic SWNTs of various lengths exposed to air. The data (squares) was obtained in the course of this work and from Refs. [14, 23]. The dashed trend line is a least-squares fit using Eq. 3, obtaining a 5 V/μm slope. SWNT breakdown occurs by oxidation when the peak temperature (in their middle) reaches $T_{BD} \approx 600$ °C [18]. Inspection by AFM imaging revealed that tubes were indeed broken at or very near their midpoint, as expected [14].

**Figure 3:** Measured electrical *I-V* characteristics up to breakdown for a 3 μm long metallic SWNT (symbols, from Fig. 4c of Ref. [14]) and simulation results using the model described in the text (solid line). The simulation is stopped once the peak SWNT temperature (in the middle of the tube) reaches $T_{BD}$ = 873 K, which occurs at $I_{BD} \approx 17.2$ μA and $V_{BD} \approx 15.2$ V here. The inset shows the computed temperature profile along the SWNT at 3, 6, 9, 12 and 15 V bias.

**Figure 4:** Sensitivity of the model to a few key parameters. The original case (black solid line) is the same as that in Fig. 3, with $R_C$ = 100 kΩ, $L$ = 3 μm, $d$ = 2 nm, $g$ = 0.17 WK$^{-1}$m$^{-1}$, and the rest as described in the text. For the three variations we increased the spontaneous OP emission length to $\lambda_{OP,300}$ = 20 nm (from 15 nm default), decreased the OP energy to $\hbar\omega_{OP}$ = 0.16 eV (from 0.17 eV), and boosted the heat dissipation per unit length into the substrate to $g$ = 0.3 Wm$^{-1}$K$^{-1}$ (from 0.17 Wm$^{-1}$K$^{-1}$).

**Figure 5:** *I-V* characteristics computed with (solid lines) and without (dashed lines) taking self-heating into account in the present model. The ambient temperature is $T_0$ = 293 K. The longest tubes show less self-heating effects at the same voltage due to lower power density and better heat sinking into the substrate. However, tubes shorter than about 1 μm are expected to benefit from more heat dissipation and sinking into their contacts.

**Figure 6: (a)** Temperature dependence of the *low-bias* resistance for a 3 μm long metallic SWNT. Symbols are experimental data, while lines represent our model with OP absorption (solid line) and without (dashed line). The subtle importance of this scattering process is evident even at low bias for ambient temperatures greater than 250 K. The electrical contact resistance was estimated to be $R_C \approx 24$ k$\Omega$ here (Pt bottom contacts, as drawn in Fig. 1). **(b)** Computed temperature dependence of the *low-bias* resistance assuming ideal electrical contacts ($R_C = 0$) for metallic SWNTs of various lengths. The quantum contact resistance ($h/4q^2 \approx 6.5$ k$\Omega$) is naturally still present (also see Fig. 7).

**Figure 7: (a)** Estimate of the various electron mean free paths (MFPs) at *low-bias* as a function of temperature ($F = 200$ V/cm, or 60 mV bias across 3 µm tube). The total MFP ($\lambda_{eff}$ in Eq. 8) is also plotted (solid line). Acoustic phonon (AC) scattering dominates at low bias, but optical phonon (OP) scattering (with MFPs $\approx 10$ µm at 300 K) must be included as the temperature rises, especially for longer tubes (also see Fig. 6). The strong decrease of the OP scattering length is owed in particular to the exponential dependence of the OP absorption process on temperature, as in Eq. 11. **(b)** Estimate of the various electron MFPs at *high-bias* and above room temperature ($F = 20$ kV/cm, or 6 V bias across 3 µm tube). Note the OP emission MFP becomes the limiting scattering mechanism, on the order of tens of nm at high temperature and high bias. The OP absorption mechanism continues its strong decrease with temperature, reaching $\lambda_{OP,abs} \approx 150$ nm near the burning point.

**Figure 8:** Comparison of thermal conductivity owed to phonons (Eq. 16 and Ref. [11]) with that due to electrons estimated from the Wiedemann-Franz Law. The low- and high-bias electron thermal conductivity corresponds to 60 mV and 6 V across a 3 µm long tube, respectively, same as the scenarios in Figs. 7a and 7b.

**Figure 9:** Order of magnitude estimates for the resistance parameters of a metallic SWNT with $L \approx 2$ µm and $d \approx 2$ nm, at room temperature. Arrows indicate estimates obtained from the Wiedemann-Franz Law showing the heat flow resistance (conductance) owed to electrons alone is significantly larger (smaller). Hence, phonons dominate heat conduction even along metallic nanotubes (and nanotube contacts) at all temperatures of practical interest [48].

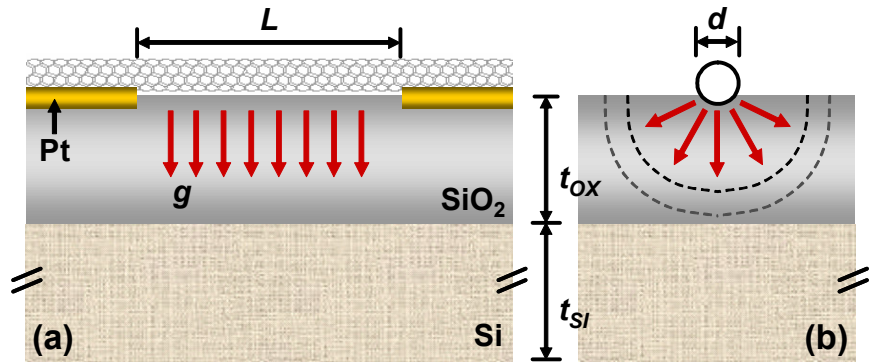

**Figure 1.**

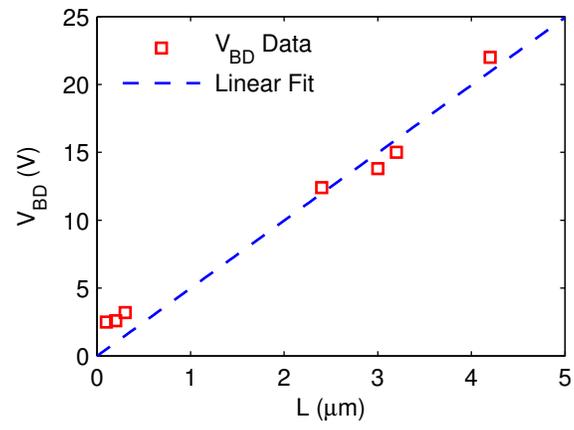

**Figure 2.**

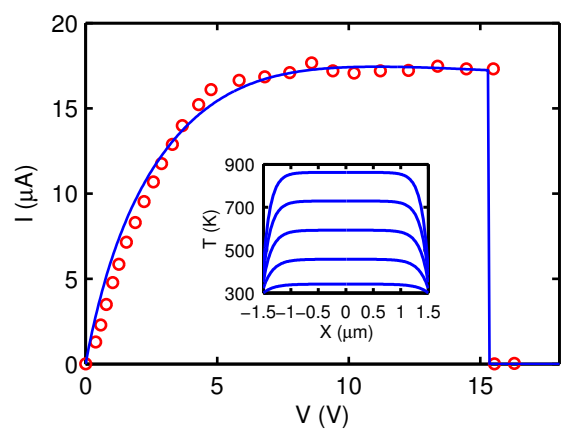

**Figure 3.**

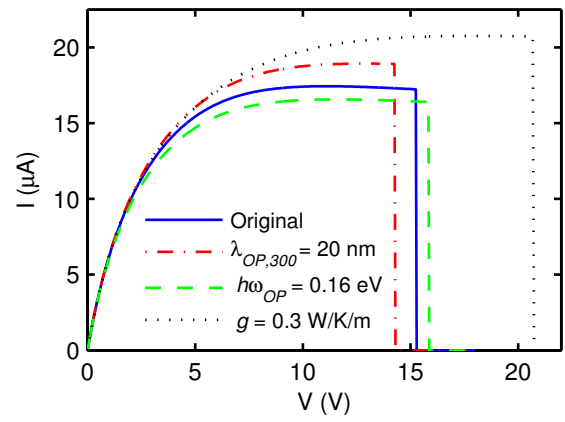

**Figure 4.**

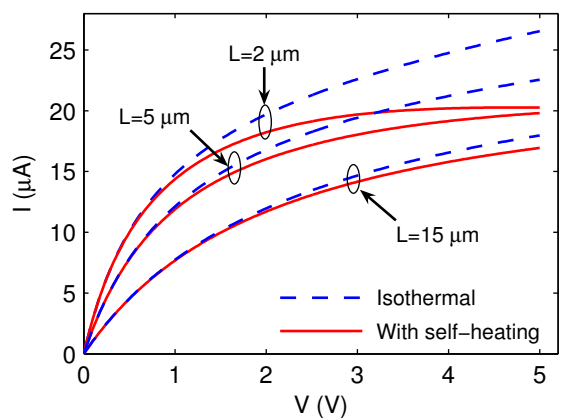

**Figure 5.**

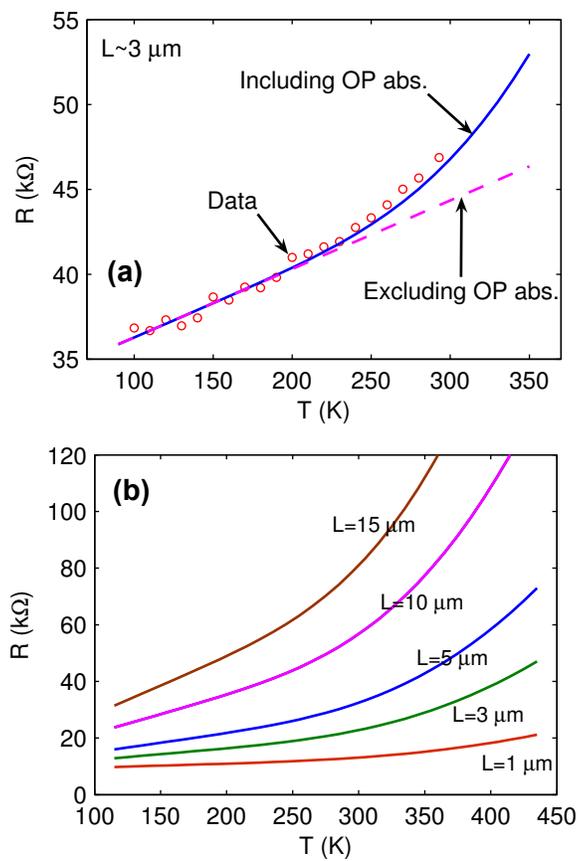

**Figure 6.**

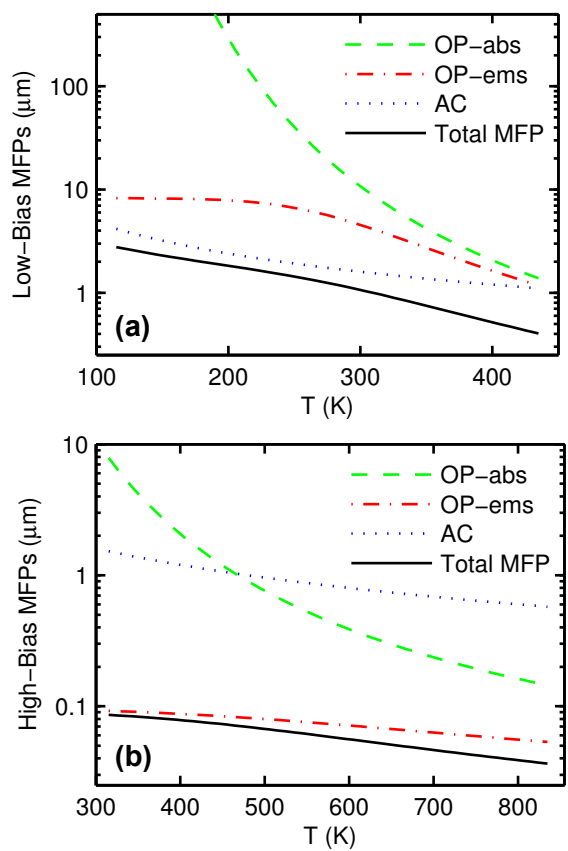

**Figure 7.**

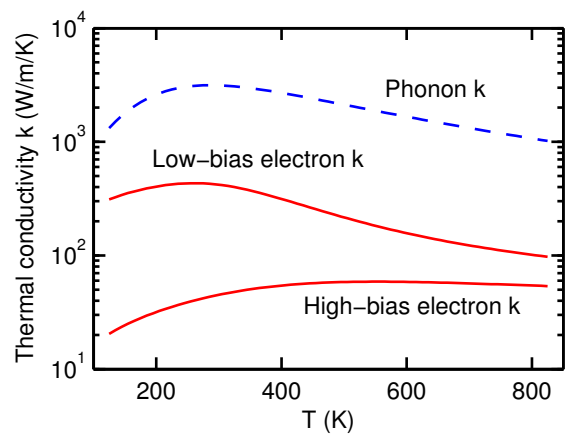

**Figure 8.**

|  | Thermal Resis. Phonons (K/W) | Thermal Resis. Electrons (K/W) | Electrical Resis. Electrons (kΩ) |
|---|---|---|---|
| Nanotube | 3 x 10$^8$ | 2 x 10$^9$ ← | ~ 15 |
| Contacts | 5 x 10$^6$ | 2 x 10$^9$ ← | ~ 15 |

**Figure 9.**